\documentclass[
 reprint,
 superscriptaddress,
 preprintnumbers,
 nofootinbib,
 amsmath,amssymb,
 aps,
 prl,
]{revtex4-2}
\usepackage{psfrag,slashed,cancel,array,graphicx,hyperref}
\usepackage{subfigure}
\usepackage[labelfont=bf,labelsep=quad,justification=justified]{caption}
\usepackage[utf8]{inputenc}
\usepackage{mathtools}
\usepackage{mathrsfs}
\usepackage{multirow}
\usepackage{braket}
\usepackage{slashed}
\usepackage{booktabs}
\usepackage[normalem]{ulem}
\usepackage{slashed}
\usepackage{simpler-wick}
\hypersetup{pdftitle={A Conformal Bridge for the Light Transform of QCD Correlation Functions},pdfcreator={},linkcolor=[rgb]{0.15,0.35,0.75},colorlinks=true,citecolor=[rgb]{0.675,0,0.2},urlcolor=[rgb]{0.15,0.35,0.65}}

\def\beq{\begin{equation}}
\def\eeq{\end{equation}}
\def\bsp#1\esp{\begin{split}#1\end{split}}

\newcommand{\refcite}[1]{ref.~\cite{#1}}

\newcommand{\eq}[1]{eq.~\eqref{eq:#1}}

\newcommand{\cL}{\mathcal{L}}

\newcommand{\cO}{\mathcal{O}}

\newcommand{\nn}{\nonumber}

\newcommand{\df}{\mathrm{d}}

\newcommand{\bq}{{\bar q}}

\newcommand{\as}{\alpha_s}
\newcommand{\bt}{{\vec b}_T}
\newcommand{\qt}{{\vec q}_T}

\newcommand{\orange}[1]{{\color{orange}{#1}}}

\newcommand{\JQQC}{J^{\rm QQC}}
\AtBeginDocument{
\heavyrulewidth=.08em
\lightrulewidth=.05em
\cmidrulewidth=.03em
\belowrulesep=.65ex
\belowbottomsep=0pt
\aboverulesep=.4ex
\abovetopsep=0pt
\cmidrulesep=\doublerulesep
\cmidrulekern=.5em
\defaultaddspace=.5em
}

\def\cO{\mathcal{O}}
\def\cC{\mathcal{C}}

\def\zb{\bar{z}}

\def\eps{\epsilon}

\def\be{\begin{equation}}
\def\ee{\end{equation}}

\begin{document}

\preprint{MIT-CTP 6016, CERN-TH-2026-005}

\title{A Conformal Bridge for the Light Transform of QCD Correlation Functions}

\author{Hao Chen}
\email{hao\_chen@mit.edu}
\affiliation{Center for Theoretical Physics, Massachusetts Institute of Technology, Cambridge, MA 02139, USA}
\author{Pier Francesco Monni}
\email{pier.monni@cern.ch}
\affiliation{CERN, Theoretical Physics Department, CH-1211 Geneva 23, Switzerland}
\author{Zhaoyan Pang}
\email{zypang@stu.pku.edu.cn}
\affiliation{School of Physics, Peking University, Beijing 100871, China}
\author{Gherardo Vita}
\email{gherardo.vita@cern.ch}
\affiliation{CERN, Theoretical Physics Department, CH-1211 Geneva 23, Switzerland}
\author{Hua Xing Zhu}
\email{zhuhx@pku.edu.cn}
\affiliation{School of Physics, Peking University, Beijing 100871, China}
\affiliation{Center for High Energy Physics, Peking University, Beijing 100871, China}

\begin{abstract}
Understanding the link between correlation functions (CFs) of local operators and measurable collider correlators has emerged as a new opportunity in the study of gauge theory dynamics at colliders.
While in Conformal Field Theories (CFTs) this connection is established by the light transform, the non-conformal nature of QCD complicates its use beyond the lowest perturbative order.
We show that a continuation of the CFs to the Wilson-Fisher fixed point can be used as a method to overcome these obstacles, serving as a \emph{conformal bridge} for the evaluation of the light transform.
At the fixed point, the renormalized CF of four local operators features a variable drop and only depends on two conformal cross ratios, in line with a genuine CFT quantity. This allows us to exploit CFT techniques to perform, for the first time, its light transform at higher loop orders. Remarkably, the resulting collider correlator in four dimensions can be recovered from this result simply by using lower-loop data.
We demonstrate this method by computing the back-to-back limit of the charge-charge correlation (QQC) at two loops in QCD through the light transform of the CF of four vector currents in the sequential light-cone limit, reproducing a recent prediction.
\end{abstract}

\maketitle

\paragraph*{Introduction.---}

A central challenge in quantum field theory is to bridge the gap between fundamental correlation functions and experimentally measurable observables. Collider correlators \cite{Basham:1978bw,Basham:1978zq,Donoghue:1979vi,Ellis:1980wv,Kunszt:1992tn,Belitsky:2001ij} are standard probes of the dynamics of quarks and gluons in Quantum Chromodynamics (QCD) and play a central role in the study of the strong interaction in collider physics. While these observables are often computed via scattering amplitudes techniques, an alternative formulation based on local correlation functions offers a more direct link to the underlying field theory and the potential to expose hidden analytic structure.

This connection was originally developed in the context of maximally supersymmetric conformal field theories (CFTs) \cite{Hofman:2008ar,Belitsky:2013xxa,Belitsky:2013bja}, and employs techniques such as Mellin representation of correlation functions \cite{Symanzik:1972gau,Mansouri:1973aa,Dobrev:1975ru,Arutyunov:2000ima,Penedones:2010ue,Fitzpatrick:2011ia,Paulos:2011ie} and double discontinuities of four-point functions \cite{Belitsky:2013xxa,Belitsky:2013bja,Henn:2019gkr,Chicherin:2020azt}. These tools are deeply rooted in conformal symmetry, relying on the fact that correlators in a CFT depend only on a limited set of cross-ratios, which drastically simplifies the analytic treatment of detector observables.

Remarkably, massless QCD inherits a form of emergent conformal symmetry at lowest order in perturbation theory, since the beta function starts at ${\cal O}(\alpha_s^2)$ in $4$ dimensions. This allows conformal techniques to be exploited even in a theory that is not truly conformal, as it was done in the case of the four-point correlation functions \cite{Chicherin:2020azt}. However, beyond tree level, the running of the coupling introduces a dependence on an unphysical renormalization scale and new kinematic structures emerge as a result of the absence of conformal symmetry, which obstruct the application of the CFT-inspired analytic machinery.

\begin{figure}
  \centering
  \includegraphics[width=0.5\textwidth]{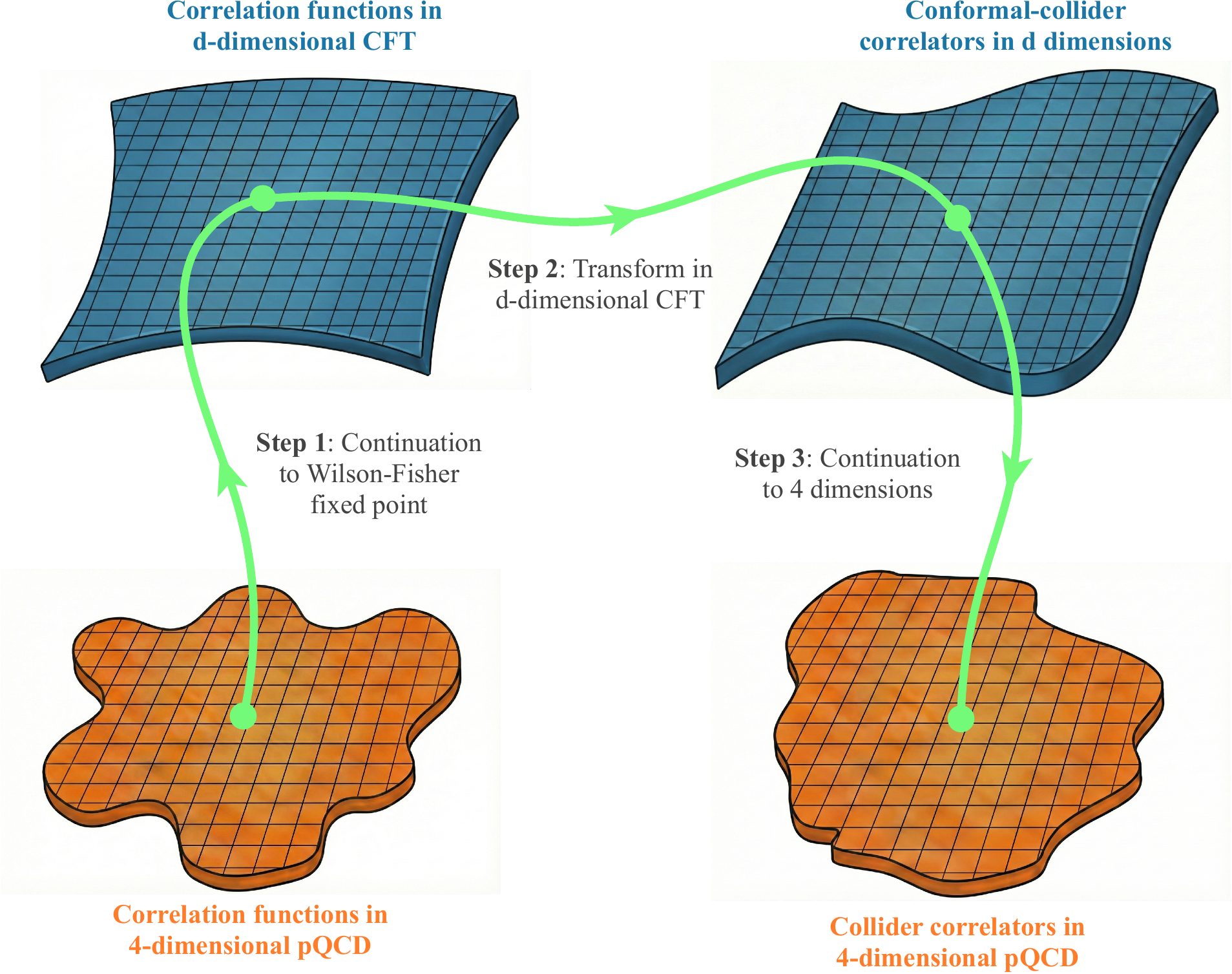}
  \caption{Illustration of the conformal bridge procedure.}
  \label{fig:bridge}
\end{figure}

In this Letter, we propose a method to circumvent this obstruction. By working at the conformal, Wilson-Fisher~\cite{Wilson:1971dc,Wilson:1973jj} point of QCD, namely the space-time dimension $d=4-2\,\epsilon$ at which the beta function
\beq\label{eq:beta}
\beta[\as, \epsilon] = -2 \as\left[ \epsilon + \frac{\as}{4\pi} \beta_0 + \left(\frac{\as}{4\pi}\right)^2 \beta_1 + \dots \right]
\eeq
vanishes, we show that the CFs retain their conformal symmetry even beyond the lowest perturbative order.
This observation allows us to apply CFT techniques to evaluate the light transform of the CF to obtain the corresponding conformal collider correlator. Crucially, the latter can be continued back to four dimensions using solely information from lower perturbative orders. A schematic illustration of the procedure is depicted in fig.~\ref{fig:bridge}.
This approach enables the treatment of QCD correlators using CFT techniques, and opens a path toward computing collider correlators beyond $\cO(\alpha_s)$ directly from correlation functions bypassing standard amplitude methods. Prior to this work, the Wilson-Fisher fixed point has found several other applications in QCD (cf.,~e.g., Refs.~\cite{Broadhurst:1993ru,Braun:2003rp,Vladimirov:2016dll,Braun:2018mxm,Duhr:2022yyp,Moult:2022xzt,Balitsky:2024xvi,Brunello:2025rhh}).

We demonstrate this method by calculating the back-to-back limit of the charge-charge correlation, reproducing a recent prediction from Soft-Collinear Effective Theory \cite{Monni:2025zyv}.
This result also allows us to shed more light onto the connection between asymptotic limits of CFs and the ones of collider correlators in QCD.

\paragraph*{From CFs to collider correlators.---}
Local correlation functions are closely related to collider correlators by the detector limit and detector integral~\cite{Belitsky:2013xxa,Belitsky:2013bja,Belitsky:2013ofa,Chicherin:2020azt}. The starting point of our study is the CF of four electromagnetic currents in QCD, $J^\mu(x_i) = \bar{\psi}(x_i)\gamma^\mu\psi(x_i)$, where, without loss of generality, we considered all quarks to have unit electric charge. The conclusions will hold for physical charges as well. The CF can be expressed in terms of the following Wightman expectation value
\be~\label{eq:JJJJ}
   G^{\mu\nu\rho\sigma}(\{x\}) =   \langle J^\mu(x_1) J^\nu(x_2) J^\rho(x_3) J^\sigma(x_4) \rangle_W\,,
\ee
where the expectation value $\langle \cdot \rangle_W$ is taken in the interacting vacuum of the theory. We start from the Euclidean case in which the points $x_i^\mu$ are space-like separated, and we will perform an analytic continuation to the Minkowski region later.
The CF in eq.~\eqref{eq:JJJJ} is related to the charge-charge correlation (QQC) in $e^+ e^-$ collisions at c.o.m.~energy $\sqrt{s}$. Denoting the angular separation between the two charge detectors by
\be
z = \frac{q^2 (n \cdot n^\prime)}{2 (n\cdot q)(n^\prime \cdot q)}\,,\quad q^\mu = \sqrt{s}\, (1,\vec{0})\,,
\ee
the QQC is obtained via the following relation~\cite{Chicherin:2020azt}
\begin{align}\label{eq:QQC}
    \mathrm{QQC}(z)=&\int d^2 \vec{n} \,d^2 \vec{n}'\delta\left(z-\frac{q^2 (n \cdot n^\prime)}{2 (n\cdot q)(n^\prime \cdot q)}\right)\\&\times \int d^4 x\, e^{i q\cdot x} \langle J^\mu(x) {\cal Q}(\vec{n}) {\cal Q}(\vec{n}^\prime) J_\mu(0) \rangle_W\notag\,,
\end{align}
where $n^\mu = (1,\vec{n})$, $n^{\prime\,\mu} = (1,\vec{n}^\prime)$.
The charge detector operator ${\cal Q}(\vec{n})$ is defined as the light transform $\mathbb{L}_{n,x}^\mu$ of the vector current $J^\mu(x)$, with
\begin{equation}\label{eq:LT4}
  \mathbb{L}^\mu_{n,x} = \lim_{\bar{n} \cdot x\to \infty} \frac{\bar{n}^{\mu}}{2}\int_{-\infty}^\infty \!\!\!\!d(n\cdot x)
     \left(\frac{\bar{n} \cdot x}{2}\right)^{2}  \,,
\end{equation}
where $\bar{n}^\mu$ is an auxiliary light-like vector, for instance $\bar{n}^\mu \equiv (1,-\vec{n})$. From eq.~\eqref{eq:QQC} we can relate the QQC to the CF~\eqref{eq:JJJJ} by recasting
\begin{align}\label{eq:GtoQQC}
    \!\!\!\langle J^\mu(x) {\cal Q}(\vec{n}) {\cal Q}(\vec{n}^\prime) J_\mu(0) \rangle_W =\mathbb{L}^\nu_{n,x_2}  \mathbb{L}^\sigma_{n^\prime,x_4} G_{\mu\nu\rho\sigma}(\{x\}) g^{\mu\rho},
\end{align}
where one performs the analytic continuation of $G_{\mu\nu\rho\sigma}(\{x\})$ using the Wightman prescription before taking the light transform of $J^\nu(x_{2})$ and $J^\sigma(x_{4})$~\cite{Belitsky:2013xxa,Chicherin:2020azt}.
In perturbative QCD, the charge-charge correlation diverges starting at the second perturbative order, but eq.~\eqref{eq:QQC} holds non-perturbatively at the hadron level.
Recently, it was shown that the infrared safety is restored in the back-to-back limit ($z\to 1$)~\cite{Monni:2025zyv}, and the charge-charge correlation obeys a factorization formula in terms of perturbatively calculable quantities.

Starting from eq.~\eqref{eq:QQC}, the back-to-back limit of the QQC can be related to the detector limit of the CF~\eqref{eq:JJJJ}~\cite{Korchemsky:2019nzm,Chen:2023wah} in which $x_2^\mu$ and $x_4^\mu$ are taken to future light-like infinity. In turn, in a CFT, this limit is related~\cite{Korchemsky:2019nzm} to the sequential-light-cone (SLC) limit of the CF, in which $|x_{24}^2| \sim |x_{13}^2| \gg |(x_{i+1}-x_i)^2|$. However, as we will discuss shortly, the latter correspondence becomes more subtle in a non-conformal theory like QCD, and some additional care is required.

An effective-theory description of the SLC limit of the CF~\eqref{eq:JJJJ} in QCD has been recently established in ref.~\cite{Chen:2025ffl}.
In this limit, we can write the CF as~\cite{Chen:2025ffl}
\begin{equation}\label{eq:Rdef}
    G^{\mu\nu\rho\sigma}(\{x\}) = G_{\rm tree}^{\mu\nu\rho\sigma}(\{x\}) \, R(u,v, \vec{a})\,,
\end{equation}
where $G_{\rm tree}^{\mu\nu\rho\sigma}(\{x\})$ is the tree-level perturbative prediction and we introduced the conformal cross ratios $u= x_{12}^2 x_{34}^2/(x_{13}^2x_{24}^2)$, $v = x_{23}^2 x_{14}^2/(x_{13}^2x_{24}^2)$,
with $x_{ij}^2 \equiv (x_i-x_j)^2$,
as well as the vector of additional variables
\begin{equation}
    \vec{a}\equiv (x_{13}^2 \mu^2, x_{24}^2 \mu^2, x_{12}^2 /x_{34}^2, x_{23}^2/x_{14}^2)\,.
\end{equation}
The function $R$ in eq.~\eqref{eq:Rdef} admits a perturbative expansion $R = 1+ \sum_{n=1}^\infty \bar{\alpha}_s(\mu) R^{(n)}$, with $\bar{\alpha}_s \equiv \alpha_s/(4\pi)$.
As discussed above, we work in the Euclidean region where $x^2_{ij}<0$, as other regions can be accessed via analytic continuation. The SLC limit is defined by $u,v\to 0$, while all the variables in $\vec{a}$ are of ${\cal O}(1)$. The two-loop expression for $R$ was obtained in ref.~\cite{Chen:2025ffl} and we report it in~\cite{supplemental}.

The correspondence between the SLC limit and the back-to-back limit, valid in a CFT, makes it tempting to argue that an analogous link holds in QCD.
Nevertheless, as we attempt to extract the QQC from the correlation function using~\eq{QQC}, a problem arises.
In the light-transform of the CF, the detector limit drives $ |x_{24}^2| \to \infty$, producing a logarithmic divergence in the correlation function, as it can be seen easily from the expressions reported in ref.~\cite{supplemental}.
Physically, this reflects the fact that, in this limit, the CF is sensitive to regions where $|x_{24}^2| \gg |x_{13}^2|$, where the power counting assumed to derive the SLC description of the CF is violated. These regions do not appear in a CFT due to conformal symmetry. The simplest way of seeing this is that, in this case, $R$ is a function only of the cross ratios $u,v$, which are formally left untouched by the detector limit.
Therefore, the absence of conformal symmetry in QCD prevents one from directly evaluating the light transform~\eqref{eq:LT4} to connect the SLC limit of the CF to the back-to-back limit of the QQC.
However, we will shortly show how the continuation to the Wilson-Fisher fixed point provides an elegant solution to this problem.

It is important to clarify that the divergence we encounter in the light transform is unrelated to the infrared unsafety of the QQC that starts at next-to-leading power in the back-to-back limit~\cite{Monni:2025zyv}. The latter is due to a soft gluon splitting into quarks which are then measured by the two detectors. This requires the two detectors at $x_2^\mu$ and $x_4^\mu$ to be directly connected by a fermionic line,
which contributes at sub-leading power in the SLC limit~\cite{Chen:2025ffl}. On the other hand, the divergence of the light transform appears already at leading power and would also be present in the computation of infrared safe quantities such as the energy-energy correlation.

\paragraph*{A conformal bridge from the Wilson-Fisher fixed point.---}
\begin{figure}
  \centering
  \includegraphics[width=0.3\textwidth]{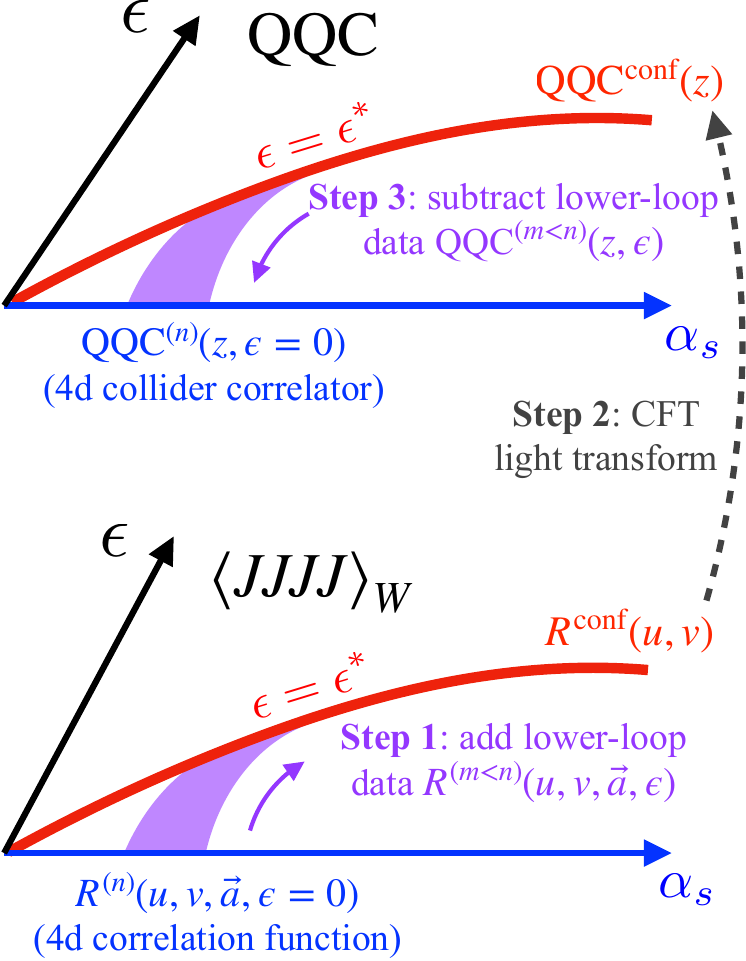}
  \caption{Application of the conformal bridge procedure to the light transform of the correlation function.}
  \label{fig:bridgeCF}
\end{figure}
From the above discussion it becomes clear that in a conformal field theory the issue is avoided altogether, since the correlation function depends only on the conformal cross ratios $u, v$.
To approach this problem in QCD, we therefore consider the theory in $d = 4 - 2\epsilon$ dimensions, where the $\beta$ function takes the form in eq.~\eqref{eq:beta}.

In this framework, there exists a critical value $\epsilon^*$ such that $\beta[\as, \epsilon^*] = 0$. At this point, known as the Wilson–Fisher fixed point~\cite{Wilson:1971dc,Wilson:1973jj}, the coupling does not run and the conformal symmetry is effectively restored at the perturbative level~\cite{Braun:2018mxm}.
Therefore, although QCD in four dimensions is not conformal, the existence of this fixed point suggests that, near $\epsilon^*$, one may attempt to extend conformal field theory techniques to QCD and explore whether its correlation functions exhibit an analogous dependence only on the conformal cross ratios.
\paragraph*{Step 1: The Sequential Light Cone Limit of CF in $d=4-2\eps$ dimensions.---}
To explore this direction, we extend the factorization theorem of \refcite{Chen:2025ffl} to $d$ dimensions, obtaining
\begin{align}\label{eq:ddimJJJJfact}
    R(u,v,\vec{a},\epsilon) = &\,{\cal W}(\{x\},\epsilon) \prod_{i=1}^4 \int_0^\infty \!\!\! \df \omega_i \,\left(\frac{y_i^-}{2\pi \omega_i}\right)^\epsilon\,e^{-\omega_i \frac{y_i^2}{2 y_i^-}}\nn \\\times &\,\cC_i(\omega_i \omega_{i+1},\epsilon) {\cal J}_{n_i}\left(\frac{\omega_i}{y_i^-},\epsilon\right)\,.
\end{align}
As discussed in \refcite{Chen:2025ffl}, ${\cal J}_{n_i}$ is a jet function that describes the propagation of energetic modes between the vertices of the CF, $\cC_i$ is a Wilson coefficient, and ${\cal W}$ is a polygonal Wilson loop.
Each object in \eq{ddimJJJJfact} is a renormalized quantity containing higher orders in $\epsilon$; they are separately smooth and finite in the $\epsilon \to 0$ limit.
Following \refcite{Chen:2025ffl}, we evaluate \eq{ddimJJJJfact} at $\epsilon=\epsilon^*$ up to $\cO(\alpha_s^2)$ and obtain (cf.~ref.~\cite{supplemental})
\begin{widetext}
{\footnotesize
\begin{align}\label{eq:variable_drop_R}
R(u,v,\vec{a},\epsilon^*) &\equiv R^\text{conf}(u,v) = 1 + \frac{\alpha_s}{4\pi}\ {\color{blue}C_F} \left(-2 L_u L_v+L_{uv}-4 \zeta_2\right)\nn\\&+\left(\frac{\alpha_s}{4\pi}\right)^2\Bigg\{ {\color{blue}C_F^2} \left[ 16 - 28 \zeta_2 - 48 \zeta_3 + 116 \zeta_4 + \left( 4 \zeta_2 - 24 \zeta_3 + \frac{5}{2} \right) L_{uv} + \left( 4 \zeta_2 - \frac{7}{2} \right) L_{uv}^2 + L_u L_v \left( 8 - 2 L_{uv} \right) + 2 L_u^2 L_v^2 \right] \nn\\
& + {\color{blue}C_F C_A}\left[ -52 - \frac{196 \zeta_2}{9} + \frac{248 \zeta_3}{3} - 16 \zeta_4 + \left( -8 \zeta_2 + 12 \zeta_3 + \frac{221}{18} \right) L_{uv} + \left( 4 \zeta_2 - \frac{134}{9} \right) L_u L_v \right] \nn\\
& + {\color{blue}C_F n_f} \left[ 8 + \frac{40 \zeta_2}{9} - \frac{32 \zeta_3}{3} - \frac{13}{9} L_{uv} + \frac{20}{9} L_u L_v \right]\Bigg\} + {\cal O}\left(\alpha^3_s\right),
\end{align}
\par}
\end{widetext}
where $L_y \equiv \ln (y)$.
Crucially, the result is independent of the four variables in $\vec{a}$. Therefore, at the conformal fixed point, the renormalized CF in QCD features a \emph{variable drop} and only depends on the conformal cross ratios $u$ and $v$. This is in line with what is expected from a quantity in a CFT. The variable drop in \eqref{eq:variable_drop_R} is a key result of this Letter.

\paragraph*{Step 2: Light transform in a $d=4-2\eps$ dimensional CFT.---}
As the correlation function now is conformal, we can apply CFT techniques to perform its light transform.
In particular, we can take the detector limit before evaluating the integral in eq.~\eqref{eq:LT4}, which is now smooth and free of divergences. As we are working away from $4$ dimensions we keep the space-time dimension $d$ generic. The detector limit (D.L.) greatly simplifies the structure of the local correlation function, giving
\begin{align}
\langle J_1 J_2 J_4 J_3\rangle_d^{\text{conf}}(u,v) \underset{\rm D.L.}{\to}\frac{F_d(u,v,\gamma)}{(n_2\cdot n_4)^{(d-1)} (-x_{13}^2)^{d}}\,,
\end{align}
where
$\gamma\equiv\frac{2(n_2\cdot x_{13})(n_4\cdot x_{13})}{(n_2\cdot n_4)x_{13}^2}$
encodes the leftover dependence on the relative angle between the light-like axes $n_2$, $n_4$ and $x_{13}^\mu$ (the sources separation)
and $F$ encodes the information on $R$ after the detector limit.
We can further simplify our description by noting that the asymptotic expansion of $F$ in the back-to-back configuration is captured by the limit $\gamma \to 0$~\cite{Chen:2023wah}. The physical significance of this limit becomes apparent if we decompose $x_{13}^\mu$ along the light-cone directions $n_2$ and $n_4$:
\begin{equation}
\gamma = \frac{2}{(n_2\cdot n_4)}\frac{(n_2\cdot x_{13})(n_4\cdot x_{13})}{\Bigl[(n_2\cdot x_{13})(n_4\cdot x_{13})+{x_{13,\perp}^2}\Bigr]}  \,.
\end{equation}
Here, $\gamma \to 0$ corresponds to the regime where the transverse impact parameter is large ($x_{13,\perp}^2 \to \infty$), the position-space dual of the source's small transverse momentum ($q_T$) expansion in QCD resummation. Geometrically, this equates to the \emph{double light cone limit} where $x_1$, $x_3$ approach the light cones of $x_2$ and $x_4$ simultaneously.
After taking the detector limit, we are left with the detector integral and Fourier transform to obtain the QQC, which crucially require the analytic continuations of the correlation function. To perform the continuation, we recast the cross ratios as
\begin{align}\label{eq:uvdef}
  u = \frac{2(n_2 \cdot x_{21})(n_4 \cdot x_{43})}{-(n_2 \cdot n_4)x_{13}^2} \,,\quad
  v = \frac{2(n_2 \cdot x_{23})(n_4 \cdot x_{41})}{-(n_2 \cdot n_4)x_{13}^2}\,.
\end{align}
The Wightman prescription of operator ordering $1 < 2 < 4 < 3$ is \cite{Luscher:1974ez}
\begin{align*}
    n_2 \cdot x_{21} &\to n_2 \cdot x_{21} - i\delta \,, &
    n_2 \cdot x_{23} &\to n_2 \cdot x_{23} + i\delta \,, \\
    n_4 \cdot x_{41} &\to n_4 \cdot x_{41} - i\delta \,, &
    n_4 \cdot x_{43} &\to n_4 \cdot x_{43} + i\delta \,,
\end{align*}
from which one can directly obtain the prescription for $u$ and $v$ via their definition in~\eq{uvdef}.
However, performing such analytic continuation at the level of the correlation function is in general non-trivial.
We perform this step again by using well established CFT techniques; in particular, we can use a method proposed by Mack~\cite{Mack:2009mi}, which we report in \refcite{supplemental}.
The light transform then effectively becomes a map from logarithms of $u$ and $v$ in the correlation function $F_d(u,v,\gamma=0)$ to distributions in the collider correlator variable $\zb\equiv 1-z$
\be
\ln^m u \ln^n v ~ \underset{\rm L.T.\,+\,F.T.}{\longrightarrow} ~ \sum_{k=-1}^{m+n-1} C^{(m,n)}_k(\epsilon) \cL_k(\zb)\,,
\ee
where $\cL_n(\zb) \equiv \left[\ln^n(\zb)/\zb\right]_+$ are standard plus distributions, $\cL_{-1}(\zb) \equiv\delta(\zb)$, and  $C^{(m,n)}_k(\epsilon)$ are $\epsilon$ dependent constants (the $\epsilon$ dependence arises from the $d$-dimensional light and Fourier transforms) which can be expanded in a power series in $\epsilon$ containing zeta values.
In \refcite{supplemental}, we provide a table with the explicit substitution rules for the transform of the QQC at 2 loops.
We note that the above method can be used to perform the analytic continuation also away from the SLC limit.

With this, we complete step 2 and obtain the leading power asymptotic expansion of the QQC in the back-to-back limit at the conformal fixed point
\begin{widetext}
{\footnotesize
\begin{align}
    \mathrm{QQC}_\text{conf}(z) &=
\sigma_0 \bigg\{ -2\delta(\zb) + {{\left(\frac{\alpha_s}{4\pi}\right)}}{\color{blue} C_F} \bigg[ 8\, \cL_1(\zb) + 12\, \cL_0(\zb) + 8\left(1 + \zeta_2\right) \delta(\zb) \bigg]\\&\hspace{-1cm}+ {{\left(\frac{\alpha_s}{4\pi}\right)^2}}\Bigg\{ {\color{blue}C_F n_f}\Bigg[ -\frac{80}{9}\,\mathcal{L}_1(\bar z) -\left(\frac{4}{3}+\frac{16}{3}\zeta_2\right)\mathcal{L}_0(\bar z) -\left(\frac{56}{9}+\frac{152}{9}\zeta_2-\frac{80}{3}\zeta_3\right)\delta(\bar z) \Bigg] \nn\\ &\qquad + {\color{blue}C_F^2}\Bigg[ -16\,\mathcal{L}_3(\bar z) -72\,\mathcal{L}_2(\bar z) +\left(-104-32\zeta_2\right)\mathcal{L}_1(\bar z) +\left(-42-96\zeta_2+32\zeta_3\right)\mathcal{L}_0(\bar z) +\left(-24-100\zeta_2+192\zeta_3-192\zeta_4\right)\delta(\bar z) \Bigg] \nn\\ &\qquad + {\color{blue}C_A C_F}\Bigg[ \left(\frac{536}{9}-16\zeta_2\right)\mathcal{L}_1(\bar z) +\left(\frac{34}{3}+\frac{88}{3}\zeta_2-48\zeta_3\right)\mathcal{L}_0(\bar z) +\left(\frac{332}{9}+\frac{1076}{9}\zeta_2-\frac{872}{3}\zeta_3+32\zeta_4\right)\delta(\bar z) \Bigg] \Bigg\}
+{\cal O}(\alpha_s^3)\bigg\}\,.\nn
\end{align}
\par}
\end{widetext}

\paragraph*{Step 3: QQC in $d$-dimensions.---}
To complete the procedure we now need to perform the continuation to $d=4$. This step requires only lower order information in perturbation theory, but higher orders in the dimensional regularization parameter. In our case we therefore need the QQC at $\cO(\alpha_s \epsilon)$. For this, we extend the factorization theorem for the back-to-back asymptotics of the QQC derived in \cite{Monni:2025zyv} to $d$-dimension, obtaining
{ \begin{align} \label{eq:QQC_d_dim}
 &\mathrm{QQC}(z,\epsilon)
 = \sigma_0\,\sum_{q} \,H_{q\bar q}(Q,\mu,\epsilon)\!\! \int\df^{2-2\eps}\vec{q}_\perp  \delta\left(1-z - \frac{q^2_\perp}{Q}\right)
 \nn\\&~~~~\times
 \int\frac{\df^{2-2\eps}\vec{b}_\perp}{(2\pi)^{2-2\epsilon}} e^{i b_T \cdot q_\perp}
   \JQQC_q\Bigl(b_T, \mu, \frac{b_T Q}{\upsilon},\epsilon\Bigr)
 \nn\\&~~~~\times
   \JQQC_\bq\Bigl(b_T, \mu, \upsilon b_T Q,\epsilon \Bigr)
\,.\end{align}}
As for the factorization theorem of \eq{QQC_d_dim}, we stress that the $\epsilon\to0$ limit of each individual object in \eq{QQC_d_dim} is finite and it reproduces the ingredients of~\cite{Monni:2025zyv}. At $\cO(\alpha_s)$ the ingredients can be easily obtained by a fixed order calculation and we report them, together with the full expression at $\cO(\alpha_s \epsilon^2)$ for $\mathrm{QQC}(z,\epsilon)$, in~\cite{supplemental}.
Note that at fixed order in $\alpha_s$ and $\epsilon$, the additional $\epsilon$ dependence arising from the Fourier transforms simply converts $\ln^n(Q b_T)$ into plus distributions in $\zb$ with epsilon dependent prefactors. Using these results, the QQC at ${\cal O}(\alpha_s^2)$ in $4$-dimensional perturbative QCD can be obtained as
\begin{widetext}
{\footnotesize
    \begin{align}\label{eq:QQC2loop}
    \mathrm{QQC}^{(2)}(z) &= \mathrm{QQC}^{(2)}_\text{conf}(z) + \big[\mathrm{QQC}^{(1)}(z,\epsilon\!=\!\epsilon^*)\big]_{\cO(\alpha_s^2)}
\nn \\ &\hspace{-1cm}=
   \sigma_0 \bigg\{ {\color{blue} C_A C_F} \bigg[ - \frac{88}{3}\, \cL_2(\zb)+ \left(\frac{140}{9} - \frac{8\pi^2}{3}\right) \cL_1(\zb) + \left(\frac{298}{3} - \frac{44\pi^2}{3} - 48\zeta_3\right) \cL_0(\zb)  + \left(\frac{860}{9} - \frac{56\pi^2}{27} + \frac{16\pi^4}{45} - \frac{344\zeta_3}{3}\right) \delta(\zb) \bigg]
\nn\\ &\hspace{-1cm}+\,
    {\color{blue} C_F \, n_f} \bigg[ + \frac{16}{3}\, \cL_2(\zb) - \frac{8}{9}\, \cL_1(\zb) + \left(-\frac{52}{3} + \frac{8\pi^2}{3}\right) \cL_0(\zb) + \left(-\frac{152}{9} + \frac{32\pi^2}{27} - \frac{16\zeta_3}{3}\right) \delta(\zb) \bigg]
\nn \\[6pt] &\hspace{-1cm}+\,
    {\color{blue} C_F^2} \bigg[-16\, \cL_3(\zb) - 72\, \cL_2(\zb)- \left(104 + \frac{16\pi^2}{3}\right) \cL_1(\zb) - \left(42 + 16\pi^2 - 32\zeta_3\right) \cL_0(\zb) + \left(-24 - \frac{50\pi^2}{3} - \frac{32\pi^4}{15} + 192\zeta_3\right) \delta(\zb) \bigg]\bigg\}\,,
\end{align}
\par}
\end{widetext}
where $\big[\mathrm{QQC}^{(1)}(z,\epsilon\!=\!\epsilon^*)\big]_{\cO(\alpha_s^2)}$, given in eq.~\eqref{eq:QQCzepsfixedorder} of~\cite{supplemental}, is the ${\cal O}(\alpha_s)$ correction to the QQC, evaluated at $\epsilon=\epsilon^*$ and expanded through ${\cal O}(\alpha_s^2)$.
The result in \eq{QQC2loop} exactly matches the QCD result from the SCET factorization formula derived in~\refcite{Monni:2025zyv}.
We stress that $\big[\mathrm{QQC}^{(1)}(z,\epsilon\!=\!\epsilon^*)\big]_{\cO(\alpha_s^2)}$, which is required to continue to 4 dimensions the conformal result of $\mathrm{QQC}^{(2)}_\text{conf}(z)$, is obtained exclusively from ingredients at lower orders in perturbation theory (cf.~Ref.~\cite{supplemental}).
The derivation of eq.~\eqref{eq:QQC2loop} from the light transform of a two-loop correlation function is a novel result of this work. This constitutes a non-trivial proof of the consistency between the factorization theorems in eqs.~\eqref{eq:ddimJJJJfact},~\eqref{eq:QQC_d_dim}, as well as a first explicit realization of the connection between renormalized correlation functions and collider correlators beyond the leading order in full QCD.

\paragraph*{Conclusions and outlook.---}
In this Letter we have studied the relation between CFs of local operators in QCD and the corresponding collider correlators by means of the light transform. We have shown how the continuation of the QCD correlation function of four vector currents to the Wilson-Fisher fixed point leads to a result that depends only on two conformal cross ratios, in contrast with the six variables in the four-dimensional case. This variable drop is in line with what is expected for a CFT result.
Moreover, we have exploited the conformal symmetry at the critical point to perform, for the first time, the light transform of the CF in the sequential-light-cone limit at higher perturbative orders, hence reproducing a recent prediction from soft-collinear effective theory factorization.
Our findings show that the emergent conformal symmetry of tree-level massless QCD, although broken by quantum effects, can be systematically leveraged as an intermediate organizing principle for the perturbative computation of collider observables. This approach opens a new avenue for computing higher-order collider observables directly from correlation functions, and suggests that other infrared-safe collider correlators may admit a similar treatment. Their investigation will be the subject of future work.

It will be also interesting to study the application of similar techniques to the computation of other quantities in non-conformal gauge theories. With this in mind, we observe that the restoration of the conformal symmetry, observed here for the four-point function, might be obscured by radiative corrections for other quantities such as scattering amplitudes (cf., e.g., the findings of ref.~\cite{Chicherin:2022gky}). A systematic explanation of the origin of the perturbative conformal-symmetry breaking in gauge theories remains an open question, and its investigation is an important challenge for the years to come.

\paragraph*{Acknowledgements.---}
We thank A. Zhiboedov for useful discussions.
ZYP and HXZ are supported by the National
Natural Science Foundation of China under contract No.
12425505. PM and GV wish to thank the Center for High Energy Physics of Peking University for hospitality while part of this work was carried out.
The work of PM is funded by the European Union (ERC, grant agreement No. 101044599).
Views and opinions expressed are however those of the authors only and do not necessarily reflect those of the European Union or the European Research Council Executive Agency. Neither the European Union
nor the granting authority can be held responsible for them.
HC is
supported by the U.S. Department of Energy, Office of
Science, Office of Nuclear Physics under grant Contract
Number DESC0011090.

\bibliographystyle{apsrev4-2}
\bibliography{refs}

\newpage

\onecolumngrid
\newpage
\appendix

\makeatletter
\renewcommand\@biblabel[1]{[#1S]}

\renewcommand{\theequation}{S.\arabic{equation}}
\newcommand{\logcolor}{blue}
\newcommand{\plusDcolor}{red}
\setcounter{equation}{0}
\makeatother

\section*{Supplemental material}
In this appendix, we report the technical details necessary to reproduce the results obtained in the Letter.

\section{Conformal map from the finite null polygon to the detector limit}
\label{app:conformalMap}
In this section we present an explicit example of a conformal transformation to connect the detector limit to a finite null polygon configuration~\cite{Korchemsky:2019nzm,Chen:2023wah}. Without loss of generality, we start from the null polygon:
\begin{equation}
	x_1^\mu=(x_1^0,x_1^1,x_1^2,x_1^3)=(0,1,0,0),\;
	x_2^\mu=(\sqrt{2},0,-1,0),\;
	x_3^\mu=(0,-1,0,0),\;
	x_4^\mu=(\sqrt{2},0,1,0).
\end{equation}
In the detector limit, the configuration approaches:
\begin{equation}
	x_1^\mu=(0,1,0,0),\;
	x_2^\mu=\alpha(1,0,-1,0),\;
	x_3^\mu=(0,-1,0,0),\;
	x_4^\mu=\alpha(1,0,1,0),
	\qquad \alpha\to\infty.
\end{equation}

These two configurations can be connected via a conformal transformation:
\begin{equation}
	x^\mu \;\mapsto\; x^{\prime\mu}
	=
	D_{\,2(\sqrt{2}-1)}
	\circ
	T_{(-1/2,\,0,\,0,\,0)}
	\circ
	K_{(\sqrt{2}-1,\,0,\,0,\,0)}\;x^\mu,
\end{equation}
where $K_b$ is a special conformal transformation,
\begin{equation}
	K_b:\;
	x^\mu \mapsto
	\frac{x^\mu-b^\mu x^2}{1-2\,b\!\cdot\! x+b^2x^2},
\end{equation}
followed by a translation $T_a$ and a dilation $D_\lambda$,
\begin{equation}
	T_a:\;x^\mu\mapsto x^\mu+a^\mu,\;
	D_\lambda:\; x^\mu\mapsto \lambda x^\mu.
\end{equation}

\section{Two-loop correlation function in dimensional regularization in the SLC limit}
We report the two-loop expression for the renormalized CF in the SLC limit in $d=4-2\epsilon$ dimensions used to derive the main result of the Letter.
We express the perturbative expansion of $R(u,v,\vec{a},\epsilon)$~\eqref{eq:ddimJJJJfact} as
\begin{equation}
    R(u,v,\vec{a},\epsilon) = 1 + \sum_{n=1}^{\infty} \left(\frac{\alpha_s(\mu)}{4\pi}\right)^n R_{n}(u,v,\vec{a},\epsilon)\,,
\end{equation}
where the first order reads~\cite{Chen:2025ffl}
\begin{align}\label{eq:one-loop}
    R_{1}&(u,v,\vec{a},\epsilon) = {\color{blue}C_F}  \,r_1(L_u,L_v) +\epsilon\,{\color{blue}C_F}  \Big[\frac{1}{2} \,r_1(L_u,L_v)\,\ln \frac{x_{13}^2 x_{24}^2 \mu^4}{16} \nn\\&+ r_2(L_{uv}, x) + r_2\big(L_{uv}, \frac{1}{x}\big)
+\frac{1}{4}\left(L_{13}^2+L_{24}^2\right)
+\ln (x)L_{13} L_{24}-\frac{1}{2}\left(L_v L_{13}^2+ L_u L_{24}^2\right)\nn\\
&+\frac{1}{4}\left(L_u^2+L_v^2+4 L_u L_v - 2 \left(L_u L_v^2 + L_v L_u^2\right)\right) - 8 \zeta_2 \left(L_u+L_v\right)+12+8\zeta_2-20 \zeta_3\Big]+ {\cal O}(\epsilon^2)\,,\nn
\end{align}
where
\begin{equation}
r_1(L_u,L_v)=-4 \zeta _2-2 L_u L_v + L_{uv}\,.
\end{equation}
We used the notation $L_{uv}\equiv L_u+L_v$, $\beta_0 = 11 C_A/3 - 2 n_f/3$, $x = x_{13}^2/x_{24}^2$, $L_{13}=\ln ({y_{1}^2}/{y_{3}^2})$, $L_{24}=\ln ({y_{2}^2}/{y_{4}^2})$ and
\begin{align}
    r_2(L_{uv}, x) =&\ 2 \text{Li}_3(-x) - 2 \ln(x) \text{Li}_2 (-x) - \frac{1}{4} L_{uv} \ln^2(x)
     - \frac{1}{2} (6 \zeta_2 + \ln^2(x) ) \ln\frac{(1+x)^2}{x} +\frac{1}{4}\ln^2 (x)\,.
\end{align}
The second order reads
\begin{align}\label{eq:two-loop}
     R_{2}&(u,v,\vec{a},\epsilon) = \ {\color{blue}C_F^2} \Bigg[
       -28 \zeta _2-48 \zeta _3+116 \zeta _4+L_u \left(8-2
   L_{uv}\right) L_v+2 L_u^2 L_v^2
   +
   \left(4\zeta _2-\frac{7}{2}\right)
   L_{uv}^2+\left(4 \zeta _2-24 \zeta
   _3+\frac{5}{2}\right) L_{uv}
   \Bigg]  \nn  \\& + {\color{blue} C_F C_A} \Bigg[
      \frac{68 \zeta _2}{9}
   +\frac{28 \zeta _3}{3}-16 \zeta
   _4+L_u L_v \left(4 \zeta _2-\frac{11
   L_{uv}}{6}-\frac{235}{18}\right) +\frac{11
   L_{uv}^2}{12}
    +\left(-\frac{112 \zeta _2}{3}+12 \zeta
   _3+\frac{221}{18}\right) L_{uv}
        \Bigg]
   \nn\\& + {\color{blue} C_F n_f}
    \Bigg[
       -\frac{8 \zeta _2}{9}+\frac{8 \zeta _3}{3}
     +
    L_u \left(\frac{L_{uv}}{3}+\frac{17}{9}\right)
   L_v+\left(\frac{16 \zeta
   _2}{3}-\frac{13}{9}\right)
   L_{uv}-\frac{L_{uv}^2}{6}
    \Bigg]
  + 16 {\color{blue} \left( C_F^2 - \frac{1}{2}C_F C_A\right) } \\&
   + {\color{blue} C_F} \orange{\beta_0}  \,\Big[r_2(L_{uv}, x) + r_2\big(L_{uv}, \frac{1}{x}\big)
+\frac{1}{4}\left(L_{13}^2+L_{24}^2\right)
+\ln (x)L_{13} L_{24}-\frac{1}{2}\left(L_v L_{13}^2+ L_u L_{24}^2\right)\Big] \nn\\&+ \frac{1}{2} {\color{blue} C_F}\,\orange{\beta_0} \, r_1(L_u,L_v) \ln \frac{x_{13}^2 x_{24}^2 \mu^4}{16}+ {\cal O}(\epsilon).\nn
\end{align}
It is now easy to see that, at the Wilson-Fisher fixed point
\begin{equation}
    \epsilon=\epsilon^*=-\sum_{i=1}\left(\frac{\alpha_s}{4\pi}\right)^i \beta_{i-1}\,,
\end{equation}
the CF features a variable drop and only depends on the conformal cross ratios via the logarithms $L_u$, $L_v$, in line with what discussed in the Letter.

\section{Analytic continuation of the correlation function with Wightman prescription}
\label{app:continuation}
We report here the method of Ref.~\cite{Mack:2009mi} to perform the analytic continuation of the correlation function in a CFT.
Ref.~\cite{Mack:2009mi} employs the Mellin representation of the correlation function, which, after the detector limit reads
\begin{equation}
    F_d(u,v,\gamma=0)=\sigma_0 \int \frac{dj_1 dj_2}{(2\pi i)^2} M_d(j_1,j_2) u^{j_1} v^{j_2}\,,
\end{equation}
such that the the back-to-back asymptotic of the QQC is obtained as
\begin{align}\label{eq:QQCfromMellin}
\lim_{z\to 1}   & \mathrm{QQC}_d^{\text{conf}}(z)=\int d^d x_{13} e^{i q\cdot x_{13}} \int_{-\infty}^\infty \prod_{i=2,4} d(n_i\cdot x_i) \frac{F_d(u,v,\gamma=0)}{(n_2\cdot n_4)^{(d-1)} (x_{13}^2)^{d}} =\sigma_0 \oint_C \frac{dj_1 dj_2}{(2\pi i)^2} M_d(j_1,j_2) K_d(j_1,j_2;z)\,,\nn
\end{align}
where $K_d(j_1,j_2;z)$ is the kinematic kernel for conformal QQC correlator in $d$-dimensions, generalizing the result in~\cite{Chicherin:2020azt}. The explicit expression of $K_d(j_1,j_2;z)$ is
\begin{align}
    K_d(j_1,j_2;z) =
    -\frac{\pi  (d-2) \Gamma \left(\frac{d}{2}\right)^3  \csc \left(\frac{1}{2}
   \pi  \left(d+2 \left(j_1+j_2\right)\right)\right)}{\Gamma \left(-j_1\right){}^2 \Gamma
   \left(-j_2\right){}^2 \Gamma \left(\frac{d}{2}+j_1+j_2+2\right) \Gamma
   \left(d+j_1+j_2\right)} (1-z)^{d+j_1+j_2-1}\,.
\end{align}
For the leading-power expansion, the contour $C$ is chosen to be infinitesimal circles around $j_{1,2}=-d/2$:
\begin{align}
    \left|j_1+\frac{d}{2}\right|<\delta,\quad \left|j_2+\frac{d}{2}\right|<\delta\,.
\end{align}
This procedure results in the mapping between logarithmic terms of the cross ratios and the QQC, summarized in Tab.~\ref{tab:substitution_rules}.

\begin{table}[ht]
\centering
\renewcommand{\arraystretch}{1.8}
\begin{tabular}{|c|p{12cm}|}
\hline
\textbf{CF term} & $\mathrm{QQC}(z)$ \textbf{distributions} \\
\hline
$1$ & $\displaystyle -2 \delta(\zb)$ \\
\hline
$L_u,\;L_v$ & $\displaystyle -2 \cL_0(\zb) - 4 \delta(\zb) + \epsilon \left( \frac{\pi^2}{3} - 4 \right) \delta(\zb)$ \\
\hline
$L_u^2,\;L_v^2$ & $\displaystyle -4 \cL_1(\zb) - 8 \cL_0(\zb) + \left( \frac{2\pi^2}{3} - 12 \right) \delta(\zb)$ \newline
$\displaystyle + \epsilon \left[ \left( \frac{2\pi^2}{3} - 8 \right) \cL_0(\zb) + \left( \frac{4\pi^2}{3} + 4\zeta_3 - 24 \right) \delta(\zb) \right]$ \\
\hline
$L_u L_v$ & $\displaystyle -4 \cL_1(\zb) - 8 \cL_0(\zb) - 8 \delta(\zb)$ \newline
$\displaystyle + \epsilon \left[ \left( \frac{2\pi^2}{3} - 8 \right) \cL_0(\zb) + \left( \frac{4\pi^2}{3} - 4\zeta_3 - 16 \right) \delta(\zb) \right]$ \\
\hline
$L_u^2 L_v,\; L_u L_v^2$ & $\displaystyle -6 \cL_2(\zb) - 24 \cL_1(\zb) + \left( \frac{2\pi^2}{3} - 28 \right) \cL_0(\zb) + \left( \frac{4\pi^2}{3} - 8\zeta_3 - 24 \right) \delta(\zb)$ \newline
$\displaystyle + \epsilon \left[ \left( 2\pi^2 - 24 \right) \cL_1(\zb) + \left( 4\pi^2 - 4\zeta_3 - 56 \right) \cL_0(\zb) \right.$ \newline
$\displaystyle \left. \quad + \left( 6\pi^2 - \frac{11\pi^4}{45} - 8\zeta_3 - 72 \right) \delta(\zb) \right]$ \\
\hline
$L_u^2 L_v^2$ & $\displaystyle -8 \cL_3(\zb) - 48 \cL_2(\zb) + \left( \frac{8\pi^2}{3} - 112 \right) \cL_1(\zb) + \left( \frac{16\pi^2}{3} - 32\zeta_3 - 96 \right) \cL_0(\zb)$ \newline
$\displaystyle + \left( 8\pi^2 - \frac{2\pi^4}{9} - 64\zeta_3 - 72 \right) \delta(\zb)$ \newline
$\displaystyle + \epsilon \left[ \left( 4\pi^2 - 48 \right) \cL_2(\zb) + \left( 16\pi^2 - 16\zeta_3 - 224 \right) \cL_1(\zb) \right.$ \newline
$\displaystyle \quad + \left( 24\pi^2 - \frac{44\pi^4}{45} - 32\zeta_3 - 288 \right) \cL_0(\zb)$ \newline
$\displaystyle \left. \quad + \left( 32\pi^2 - \frac{88\pi^4}{45} - 80\zeta_3 + \frac{8\pi^2}{3}\zeta_3 - 48\zeta_5 - 288 \right) \delta(\zb) \right]$ \\
\hline
\end{tabular}
\caption{Substitution rules mapping logarithmic terms $\ln^i u \ln^j v$ (denoted $L_u^i L_v^j$) to linear combinations of plus distributions $\cL_n(\zb)$ and $\delta(\zb)$.}
\label{tab:substitution_rules}
\end{table}

\section{QQC in $d$ dimensions}
\label{app:QQCddim}
In this section we will present the ingredients for the renormalized factorization theorem in $d$-dimensions of the QQC and derive the precise $\epsilon$ dependent map between the $\ln^n(Q b_T)$ terms and the plus distributions $\cL_n(\zb) \equiv \left[\frac{\ln^n(\zb)}{\zb}\right]_+$
The $d$-dimensional factorization theorem reads
\begin{align}
    \frac{\df \sigma^{\rm QQC}_{4-2\epsilon}}{\df \zb} &=\sigma_0 \int \df^{2-2\epsilon} \qt \delta\left(\zb - \frac{q_T^2}{Q^2}\right) \int \frac{\df^{2-2\epsilon} \bt}{(2\pi)^{2-2\epsilon}} e^{i \bt \cdot \qt} H(Q,\mu,\epsilon) \JQQC_q\Bigl(b_T, \mu, \frac{b_T Q}{\upsilon},\epsilon\Bigr) \JQQC_\bq\Bigl(b_T, \mu, \upsilon b_T Q,\epsilon \Bigr)  \,.
\end{align}
To $\cO(\alpha_s \epsilon)$ the ingredients read
\begin{align}
    H(Q,\mu,\epsilon) &= 1 + a_s\,C_F\Bigl(
6L - 2L^{2} + 2(-8+7\zeta_2)
+ \epsilon\bigl(
-3L^{2} + \tfrac{2}{3}L^{3} - 2L(-8+7\zeta_2)
+ \tfrac{1}{3}(-96+63\zeta_2+28\zeta_3)
\bigr)
\Bigr)\,, \nn \\
    \JQQC_q\Bigl(b_T, \mu, \frac{b_T Q}{\upsilon},\epsilon\Bigr)
    &=
    1
    + a_s C_F\Biggl\{
    L_b^{2} + L_b\,(3+4L_{\upsilon}) - 9\zeta_2 + 6
    \nonumber\\
    &\qquad\quad
    +\,\epsilon\,\Bigl[    \frac{L_b^{3}}{3}
    + L_b^{2}\Bigl(2L_{\upsilon}+\frac{3}{2}\Bigr)
    + L_b\,(6-9\zeta_2)
    + 2L_{\upsilon}\zeta_2
    - \frac{52}{3}\zeta_3   + \frac{3}{2}\zeta_2 + 12
    \Bigr]
    \Biggr\}
        \,,
\end{align}
where $L_Q\equiv \ln (Q^2/\mu^2)$, $L_b\equiv\ln(\bt^2 \mu^2/b_0^2)$,  $L_\upsilon\equiv\ln\left(b_T Q/\upsilon \right)$, and $a_s=\frac{\alpha_s(\mu)}{4\pi}$.
To evaluate the inverse Fourier transform, we set $\mu=Q,\upsilon=1$ and make use of
\begin{align}
    I_n(\zb,\eps) &= \int \df^{2-2\epsilon} \qt \delta\left(\zb - \frac{q_T^2}{Q^2}\right) \int \frac{\df^{2-2\epsilon} \bt}{(2\pi)^{2-2\epsilon}} e^{i \bt \cdot \qt} \ln^n(b_T^2 Q^2/b^2_0)\nn\\
    &= \int \df^{2-2\epsilon} \qt \delta\left(\zb - \frac{q_T^2}{Q^2}\right) \int \frac{\df^{2-2\epsilon} \bt}{(2\pi)^{2-2\epsilon}} e^{i \bt \cdot \qt} (-1)^n \frac{\partial^n}{\partial \alpha^n}\left.\left(\frac{Q^2 b_T^2}{b_0^2}\right)^{-\alpha}\right|_{\alpha=0}\,,
\end{align}
and of the $d$-dimensional FT of a power
\begin{align}
\int \frac{\df^{2-2\epsilon} \bt}{(2\pi)^{2-2\epsilon}} e^{i \bt \cdot \qt} (b_T^2)^{-\alpha} = 2^{-2\alpha}\,\pi^{\epsilon-1}\,\frac{\Gamma(1-\epsilon-\alpha)}{\Gamma(\alpha)}\,\big(q_T^2\big)^{\alpha+\epsilon-1},
\end{align}
so that \begin{align}
\label{eq:In}
    I_n(\zb,\eps) &= \int q_T^{-2\eps} \df q_T^2 \frac{\pi^{1-\eps}}{\Gamma(1-\eps)}  \delta\left(\zb - \frac{q_T^2}{Q^2}\right)  (-1)^n \frac{\partial^n}{\partial \alpha^n}\left.\left(\frac{Q^2}{b_0^2}\right)^{-\alpha}2^{-2\alpha}\,\pi^{\epsilon-1}\,\frac{\Gamma(1-\epsilon-\alpha)}{\Gamma(\alpha)}\,\big(q_T^2\big)^{\alpha+\epsilon-1}\right|_{\alpha=0}\nn\\
    &= \frac{\partial^n}{\partial \alpha^n}\left[\zb^{-1-\alpha} b_0^{-2\alpha}2^{2\alpha}\,\frac{\Gamma(1-\epsilon+\alpha)}{\Gamma(-\alpha)\Gamma(1-\eps)}\right]_{\alpha=0}\,.
\end{align}
In evaluating \eq{In}, it is important to notice that the $\alpha=0$ limit is subtle due to the behavior of $\zb^{-1-\alpha}$. In particular, we expand $\zb^{-1-\alpha}$ in distributions before taking the derivative. We have checked that the higher order poles in $\alpha$ for $n>0$ are needed to generate the terms proportional to $\delta(\zb)$ in the final expression, but do not spoil the fact that \eq{In} is finite at $\alpha=0$ for any $n$.
From this we can build the following set of substitution rules
\begin{align}
    \textcolor{\logcolor}{L_b^0} &\to \textcolor{\plusDcolor}{\delta(\zb)} \,,\nn \\
    \textcolor{\logcolor}{L_b^1} &\to -\textcolor{\plusDcolor}{\cL_0(\zb)} - \epsilon \zeta_2 \textcolor{\plusDcolor}{\delta(\zb)}\,, \nn \\
    \textcolor{\logcolor}{L_b^2} &\to 2 \textcolor{\plusDcolor}{\cL_1(\zb)} + 2\epsilon \left(\zeta_2 \textcolor{\plusDcolor}{\cL_0(\zb)}
    +\zeta_3 \textcolor{\plusDcolor}{\delta(\zb)}\right) \,,\nn \\
    \textcolor{\logcolor}{L_b^3} &\to -3 \textcolor{\plusDcolor}{\cL_2(\zb)} -4\zeta_3\textcolor{\plusDcolor}{\delta(\zb)} - 6\epsilon \left( \zeta_2 \textcolor{\plusDcolor}{\cL_1(\zb)} +\zeta_3  \textcolor{\plusDcolor}{\cL_0(\zb)} +\zeta_4 \textcolor{\plusDcolor}{\delta(\zb)}\right) \,.
\end{align}

Combining all ingredients and evaluating the inverse Fourier transform we then get to the result for the QQC to $\cO(\alpha_s \eps)$ which reads
\begin{align}\label{eq:QQCzepsfixedorder}
\mathrm{QQC}(z,\epsilon)
&=
\sigma_0 \bigg\{ -2\,\delta(\zb)
+ \frac{\alpha_s}{4\pi} C_F \Bigl[
12\,\cL_0(\zb)
+ 8 \,\cL_1(\zb)
+ 8 \bigl(1+\zeta_2\bigr)\,\delta(\zb)\Bigr]
\nonumber\\
&
+ \frac{\alpha_s}{4\pi}  C_F \epsilon \Bigl[
-12 \,\cL_1(\zb)
- 8 \,\cL_2(\zb)
- 8 \,(-3+4\zeta_2)\,\cL_0(\zb)
- 4 \,(-4+9\zeta_2-12\zeta_3)\,\delta(\zb)
\Bigr] +\cO(\alpha_s^2,\alpha_s \epsilon^2)\bigg\}
\,.
\end{align}

\end{document}